\documentstyle[aps,twocolumn,epsf,floats]{revtex}

\begin{document}
\draft
\wideabs{

\title{Incompressible states of negatively charged magneto-excitons}

\author{
   Arkadiusz W\'ojs$^{a*}$, Pawel Hawrylak$^b$, and John J. Quinn$^{a*}$}

\address{
   $^a$Department of Theoretical Physics, 
       University of Tennessee, Knoxville, Tennessee 37996-1501 \\
   $^b$Institute for Microstructural Sciences, 
       National Research Council of Canada, Ottawa, Canada K1A 0R6}

\maketitle

\begin{abstract}
   We study the system of up to four negatively charged magneto-excitons 
   ($X^-$'s) in the spherical geometry, using the exact-diagonalization 
   techniques.
   At low energies, $X^-$'s are bound and behave like charged particles 
   without internal dynamics.
   The pseudopotential describing $X^-$--$X^-$ scattering is almost 
   identical to that of electrons and the low-lying few-$X^-$ states 
   correspond to the few-electron states.
   The total angular momentum of the ground state depends on the 
   effective filling factor $\nu$ and vanishes at its special values.
   The analogs to the Laughlin $\nu=1/3$ state and the Jain $\nu=2/5$ 
   state of electrons are found.
   The $X^-$ system is predicted to exhibit the fractional quantum Hall 
   effect.
\end{abstract}

}

In a magnetic field, a pair of quasi-two-dimensional (2D) electrons and 
a valence hole can form a bound {\em negatively charged magneto-exciton} 
($X^-$) \cite{chen1,wojs1,palacios}.
This state has lower energy than the multiplicative eigenstate predicted 
by the hidden-symmetry arguments \cite{lerner}, in which exciton and 
electron do not interact with each other.
Like an electron, a bound $X^-$ is long-lived \cite{palacios}, has a mass 
and an electric charge and its lowest energy states form a degenerate Landau 
level (LL) \cite{wojs1}.
Hence, in analogy to the system of electrons, one can expect that $X^-$'s 
might form an incompressible liquid and exhibit the fractional quantum 
Hall effect (FQHE) \cite{fqhe}.
The occurrence of incompressible electron states in the thermodynamic 
limit is related to the peculiar dependence of the energy spectrum of the 
finite system on the filling factor and angular momentum 
\cite{laughlin,chen2,haldane}.
In this paper we report on the studies of the system of up to four $X^-$'s 
in the spherical geometry \cite{haldane} using the exact-diagonalization 
techniques.
We show that the low-energy states of a few-$X^-$ system are determined 
by the repulsion between $X^-$'s whose internal motion is not affected
by the $X^-$--$X^-$ scattering.
The pseudopotential describing the $X^-$--$X^-$ scattering in the lowest 
LL is very similar to that of electrons.
Consequently, the $X^-$'s are predicted to exhibit a number of properties 
characteristic of a 2D electron system in strong magnetic fields, including 
the FQHE.

We consider a system of $2N$ electrons ($e$) and $N$ valence holes ($h$), 
i.e. $N$ negatively charged excitons ($X^-$), on the Haldane sphere of 
radius $R$ \cite{haldane}.
The magnetic monopole of strength $2S$ ($2S$ is an integer) produces 
a magnetic field $B$ normal to the surface and the total flux $\Phi=4\pi 
R^2B=2S\phi_0$, where $\phi_0=hc/e$ is the flux quantum.

The single-particle (SP) states are called monopole harmonics \cite{wuyang}.
For given $2S$ they are labeled by angular momentum $l$ and its projection 
$m$.
The SP energies form degenerate LL's labeled by $l$. 
In high magnetic fields we assume that only the lowest, spin-polarized LL 
(with $l=S$) is occupied, and the SP states can be uniquely labeled by $m$.
The energy of the lowest LL is ${1\over2}\hbar\omega_c$, where $\omega_c$ 
is the cyclotron frequency, and the characteristic wavefunction length 
scale is the magnetic length, $\lambda$ ($R^2=S\lambda^2$).

The Hamiltonian of an interacting $e$--$h$ system contains the constant 
SP energy and the $e$--$e$, $e$--$h$, and $h$--$h$ Coulomb interactions.
The Hilbert space of $e$--$h$ states is spanned by SP configurations,
which are classified by total angular momentum projection $M$.
The eigenstates are obtained through diagonalization of the Coulomb
interaction in the $M$-eigensubspaces.
The eigenenergies fall into degenerate $L$-multiplets with 
$M=-L,-L+1,\dots,L$.

The ground state (GS) of an exciton $X$ ($e$--$h$ pair) is the $L=0$ state.
Due to the hidden symmetry associated with equal strength of $e$--$e$, 
$e$--$h$, and $h$--$h$ interactions (the $e$ and $h$ wavefunctions have 
the same characteristic length scale $\lambda$), the spectrum of an 
$X^-$ contains a multiplicative eigenstate ($L$-multiplet) which consists 
of an $X$ in its $L=0$ ground state decoupled from the second electron 
with $l=S$ \cite{chen1,wojs1,palacios,lerner}.
Thus, the Coulomb energy of this state is equal to the $X$ binding energy 
and its angular momentum is $L=S$.

There is {\em exactly one} state in the $X^-$ spectrum with lower energy 
than the multiplicative state; its angular momentum is $L=S-1$.
An $X^-$ in this (ground) state is a bound complex.
It has a characteristic size, an electric charge, a mass, and nonzero 
angular momentum (a degenerate $(2S-1)$-fold LL) \cite{wojs1}.
As shown by Palacios et al. \cite{palacios}, this complex is also long-lived 
(i.e. has an infinite optical recombination lifetime).

The above properties make an $X^-$ very similar to a charged particle 
without an internal structure (e.g., an electron).
Hence, one might expect that at low densities, when the charge-charge 
interaction between $X^-$'s is too weak to excite the internal degrees
of freedom, the system of $X^-$'s would behave like a system of electrons.
An interesting question is whether it can form an incompressible liquid
and exhibit the FQHE.
In order to answer it, we will compare the spectra of $NX^-$ at a given 
monopole strength $2S$, i.e. with individual angular momenta $l=S-1$,
with the spectra of $Ne$ with the same {\em individual angular momenta} 
$l$, i.e. at the monopole strength $2(S-1)$.

\begin{figure}[t]
\epsfxsize=3.25in
\epsffile{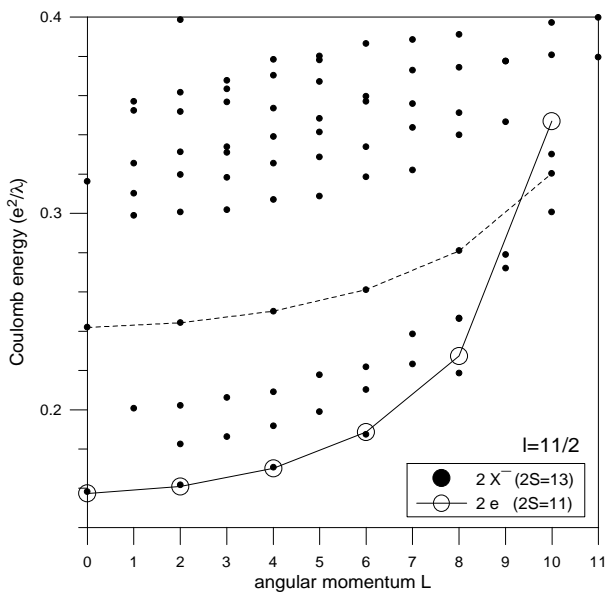}
\caption{
   Energy spectrum of two $X^-$'s with $l=11/2$ (dots) and of two 
   electrons with the same $l$ (circles).
   The $2X^-$ energies do not include GS energies of individual $X^-$'s.
   Dashed line connects the lowest multiplicative states}
\label{fig1}
\end{figure}

Let us first look at the case of $N=2$.
In Fig.~\ref{fig1}, the dots show bottom part of the energy spectrum 
of two $X^-$'s with $l=11/2$, i.e. for the monopole strength $2S=13$.
The circles show the corresponding spectrum of two electrons, the $e$--$e$ 
Coulomb pseudopotential $V_{ee}(L)$, for the same $l$, i.e. $2S=11$.
In order to compare the two spectra, two GS energies (GSE's) of a single 
$X^-$ were subtracted from the $2X^-$ energies.
For the $2X^-$ states composed of two GS $X^-$'s, this means excluding
the internal energy of the complexes and showing only the energy of
$X^-$--$X^-$ interaction, i.e. the pseudopotential $V_{X^-X^-}(L)$.

The $2X^-$ spectrum in Fig.~\ref{fig1} contains a low-lying band of states 
with even angular momenta, $L=0$, 2, 4, 6, and 8.
The energy vs. $L$ dependence within this band is virtually identical to 
that for two electrons, given by $V_{ee}(L)$.
This tells that indeed, at low energies, two $X^-$'s behave like a pair of 
electrons with the same {\em individual angular momenta} $l$, i.e., at a 
different magnetic field (also, the energy scale in Fig.~\ref{fig1}, 
$e^2/\lambda=\sqrt{S}e^2/R$, is different for $X^-$'s and for electrons).

The reason for the correspondence between the $2X^-$ and $2e$ spectra is 
that (at low enough density) the characteristic excitation energy of an 
individual $X^-$ (the $X^-$ binding energy) is higher than characteristic 
$X^-$--$X^-$ repulsion energy.
Consequently, the low-lying $2X^-$ states can be viewed as pure inter-$X^-$ 
excitations, defined by a pseudopotential $V_{X^-X^-}(L)$, while excitations 
involving internal degrees of freedom of individual $X^-$'s have higher 
energies.

A slight discrepancy between the $V_{ee}(L)$ and $V_{X^-X^-}(L)$ 
pseudopotentials appears at higher $L$'s, i.e. in the states with 
decreasing average $e$--$e$ or $X^-$--$X^-$ distance.
Clearly, when this distance is too small, the $X^-$--$X^-$ repulsion
becomes comparable to the $X^-$ binding energy, and coupling to the
internal degrees of freedom lowers the $2X^-$ eigenenergy compared to
that of $2e$.
Notice that a $2X^-$ state with $L=10$ and composed of two GS $X^-$,
corresponding to the highest-$L$ state of two electrons ($L=2l-1$), would
have too high energy to lie below the excitations of individual $X^-$'s.

We have also compared the $2X^-$ and $2e$ spectra for other values of $2S$ 
and found that there is always a low-energy band of $2X^-$ states with odd 
or even $L$'s (depending on the parity of $2S$), the $V_{X^-X^-}(L)$ and 
$V_{ee}(L)$ pseudopotentials are almost identical, and that the $L=2l-1$ 
state lies in the band of states with excited internal motion.

Lastly, as for a single $X^-$, the hidden symmetry \cite{lerner} leads 
to the set of multiplicative $2X^-$ eigenstates which are composed of 
two non-interacting excitons decoupled from two excess electrons.
The energies and angular momenta of these states are obtained by adding
two GS excitons to all possible states of two interacting electrons for 
a given $2S$.
We connected these states in Fig.~\ref{fig1} with a dashed line (the state
with maximum $L=2S-1=12$ lies outside the frame).

\begin{figure}[t]
\epsfxsize=3.25in
\epsffile{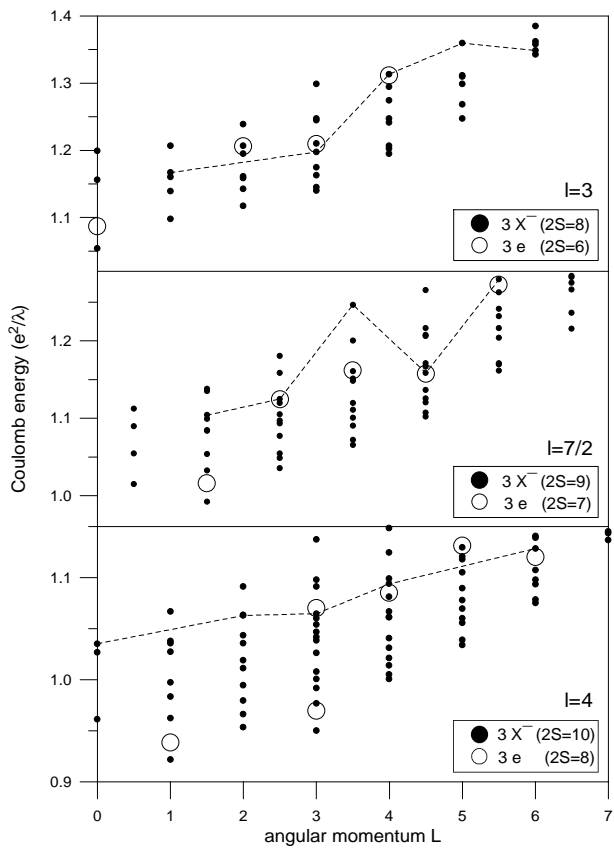}
\caption{
   Energy spectra of three $X^-$'s with $l=3$, 7/2, and 4 (dots) 
   and of three electrons with the same $l$'s (circles).
   The $3X^-$ energies do not include GS energies of individual $X^-$'s.
   Dotted line connects the lowest multiplicative states}
\label{fig2}
\end{figure}

Having shown that two $X^-$'s form states without excited internal motion,
we can now turn to the case of $N=3$.
In Fig.~\ref{fig2}, we present the energy spectra of three $X^-$'s 
with $l=3$, 7/2, and 4 (dots) and the corresponding three electron spectra 
for the same $l$'s (circles).
As for $N=2$, three $X^-$ GSE's were subtracted from the $3X^-$ energies 
and dashed lines connect lowest multiplicative states for each $L$.

As expected from the calculations for $N=2$, at the densities corresponding 
to shown values of $2S$ (exact computations for larger values of $2S$ 
require diagonalization of huge matrices), the average $X^-$--$X^-$ 
distance is rather small and hence the inter-$X^-$ excitations couple 
to the internal ones.
Nevertheless, even though the $3X^-$ energies are visibly lower than
the corresponding $3e$ energies, the angular momenta of the lowest states 
agree.
In particular, for $l=5/2$ (not shown) the low-lying state of both $3X^-$ 
and $3e$ has $L=3/2$; for $l=3$ it is $L=0$; for $l=7/2$ it is $L=3/2$;
and for $l=4$ it is $L=1$ and 3 (here also the ordering of the two 
energies agrees).

We have also carried out a similar computation for $N=4$.
For both $4X^-$ and $4e$ systems, for $l=3$ we obtained the $L=0$ GS, 
and for $l=4$ the $L=2$ GS and a low-lying $L=0$ excited state.
For $l=7/2$, there is a pair of low energy states in both spectra, with
$L=0$ and 2, however their ordering is different: the $L=0$ state is the 
GS of $4X^-$, while $L=2$ is the GS of $4e$.
We expect that the spectra of $NX^-$ and $Ne$ for $N\ge3$ become even
more similar at lower effective densities, as is in the case of $2X^-$ 
and $2e$ systems, which we were able to study in more detail.

In case of the electron system, the appearance of $L=0$ GS's is connected
with the formation of incompressible states at the special densities 
(or filling factors, $\nu=(N-1)/2S$ for the Laughlin states). 
These states are separated from the band of excitations by a gap that
is due to $e$--$e$ interactions (not due to spatial quantization) and 
persists in the thermodynamic limit.
The occurrence of this gap leads to the FQHE.
Importantly, the entire information about the $e$--$e$ interaction, which
is in turn responsible for the formation of many electron incompressible
states and the FQHE, is included in the form of the pseudopotential 
$V_{ee}(L)$ \cite{fqhe,wojs2}.

We have shown here that (at low densities) also in the system of $X^-$'s, 
the lowest sector of the energy spectrum is determined by the $X^-$--$X^-$
repulsion, and not by the internal dynamics of an individual $X^-$.
The pseudopotential $V_{X^-X^-}(L)$ is almost identical to that of 
electrons.
Consequently, at the special densities, the $X^-$'s also should form an
incompressible liquid with a nondegenerate GS (separated from the continuum 
by a gap) and exhibit the FQHE.

One last remark is noteworthy. 
In case of a finite system on a sphere, the natural definition of the 
$X^-$ filling factor for the Laughlin states is $\nu_{X^-}=(N-1)/2(S-1)$, 
and the $L=0$ GS of three $X^-$'s at $2S=8$ corresponds to the $\nu=1/3$ 
state of three electrons.
Similarly, the $L=0$ GS of four $X^-$'s at $2S=8$ corresponds to the Jain 
$\nu=2/5$ state of four electrons.

\end{document}